# Emergent Opinion Dynamics on Endogenous Networks

## László Gulyás[♥][♦], Elenna R. Dugundji[♣]

[♥]AITIA International Inc, Budapest, Hungary
[♦]Dept. of History and Philosophy of Science, Eotvos University, Budapest, Hungary
[♣]University of Amsterdam, The Netherlands
lgulyas@aitia.ai; E.R.Dugundji@uva.nl

**Abstract**

In recent years networks have gained unprecedented attention in studying a broad range of topics, among them in complex systems research. In particular, multi-agent systems have seen an increased recognition of the importance of the *interaction topology*. It is now widely recognized that emergent phenomena can be highly sensitive to the structure of the interaction network connecting the system's components, and there is a growing body of abstract network classes, whose contributions to emergent dynamics are well-understood. However, much less understanding have yet been gained about the effects of *network dynamics*, especially in cases when the emergent phenomena feeds back to and changes the underlying network topology.

Our work starts with the application of the network approach to discrete choice analysis, a standard method in econometric estimation, where the classic approach is grounded in individual choice and lacks social network influences. In this paper, we extend our earlier results by considering the endogenous dynamics of social networks. In particular, we study a model where the behavior adopted by the agents feeds back to the underlying network structure, and report results obtained by computational multi-agent based simulations.

## Introduction

In recent years networks have gained unprecedented attention in studying a broad range of topics, among them in complex systems research. In particular, multi-agent systems have seen an increased recognition of the importance of the *interaction topology*. It is now widely recognized that emergent phenomena can be highly sensitive to the structure of the interaction network connecting the system's components, and there is a growing body of abstract network classes, whose contributions to emergent dynamics are well-understood. However, much less understanding have yet been gained about the effects of *network dynamics*, especially in cases when the emergent phenomena feeds back to and changes the underlying network topology.

Our work starts with the application of the network approach to discrete choice analysis, a standard method in econometric estimation with applications, among others, in land use and transportation planning models. (Ben-Akiva 1973; Ben-Akiva and Lerman 1985; Ben-Akiva and Morikawa 1990) The field's developments over the past 30 years range from the basic random utility model to incorporate cognitive and behavioral processes, etc. However, the classic approach of discrete choice theory is fundamentally grounded in individual choice, which is in contrast with both intuition and with recent observations of the importance of social networks. Earlier work by the authors (Gulyás and Dugundji 2003a, 2003b; Dugundji and Gulyás 2004a, 2004b, 2005, 2006) considered a family of models where agents were placed on instances of abstract network classes and an agent's choice to adopt a discrete behavior was influenced by the percentages of the agent's neighbors making each choice. In this paper, we extend these results by considering the endogenous dynamics of social networks. In particular, we study a model where the behavior adopted by the agents feeds back to the underlying network structure. (E.g., a decision to commute by train instead of car increases the likelihood of train-choosing acquaintances.) We report results obtained by the application of computational multi-agent based simulations.

## Discrete Choices
## in the Presence of Social Influence

Let's consider a set $A$ of $N$ agents, indexed by $a \in [1,N]$, each of which has to choose among $M$ choices, $\Psi = \{\psi_1, \ldots, \psi_M\}$. The choices made by the agents at any fixed time $t$ are denoted by $\sigma^t_1, \ldots, \sigma^t_N$. The probabilities of the choices for the agents are given by the rule set $R$, which consists of parameterized probability distributions for each agent:

$$R = \left\{ h_{aj}(v_1,...,v_{V_{aj}}) \mid a \in [1,N], j \in [1,M] \text{ and } \sum_{j=1}^{M} h_{aj}(v_1,...,v_{V_{aj}}) = 1 \right\}$$

Here $h_{a1}()$, ..., $h_{aM}: \mathbb{R}^{V_{aj}} \to [0,1]$ are functions yielding the probability distribution for agent $a$, and $v_{aj}$, ..., $v_{V_{aj}} \in \mathbb{R}$ are parameters, $V_{aj} \in \mathbb{N}$. The social interactions are described as a set $G$ of directed links among the agents. Formally, $G \subseteq A \times A$, and the pair of (A, G) forms a directed graph. The agents take the proportion of their neighbors' choices into account. That is, it is assumed that a subset of parameters $v_{aj},...,v_{V_{aj}}$ for each agent $a$ is

$$v_{aj}^* = \frac{\sum_{g \in G(a)} \chi(\sigma_g^t = \psi_j)}{|G(a)|},$$

for each j=[1, M], where $\chi(\cdot)$ is the usual membership function.

Finally, $\Delta: G \to G$ stands for the rules of dynamic changes to graph $G$. This completes the abstract formalization of the class of models considered, which are characterized by the (N, M, R, G, Δ) quintet.

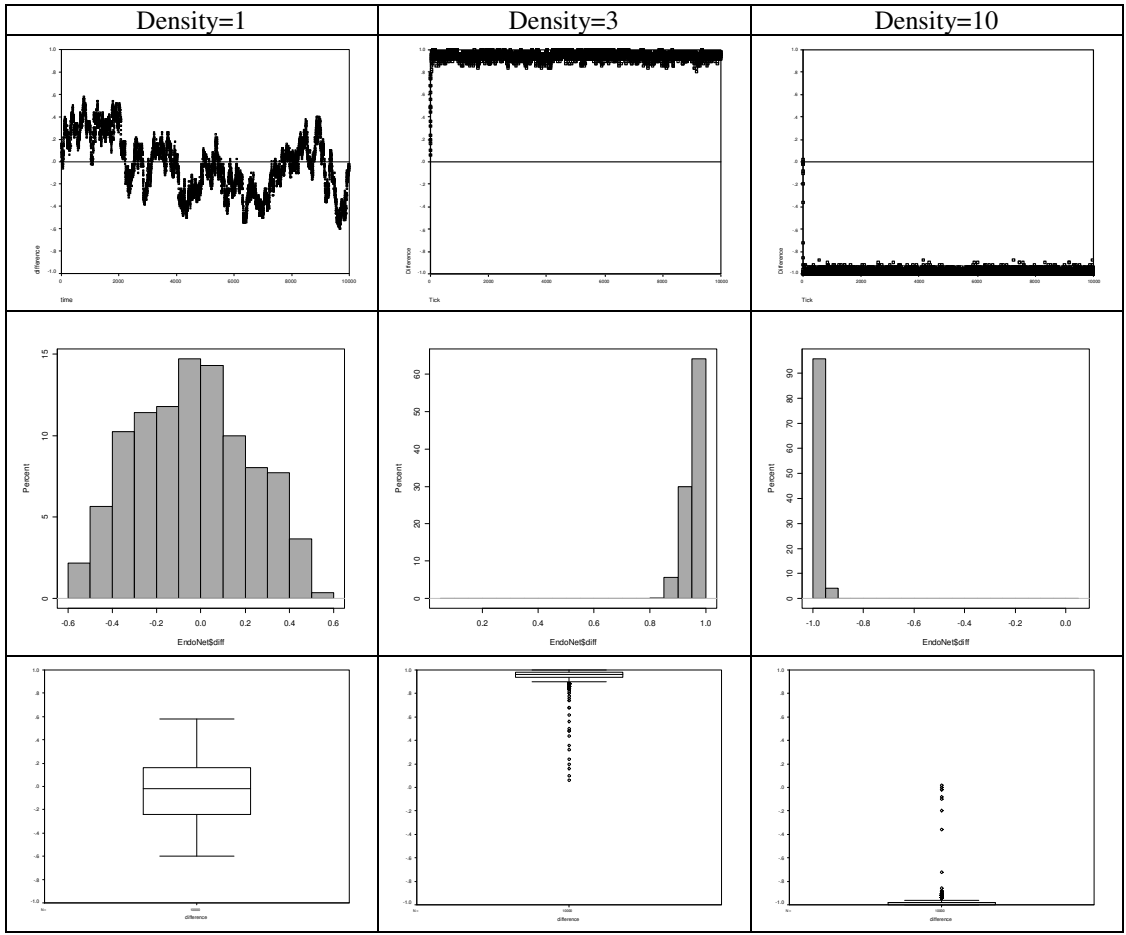

**Fig. 1: Density-dependent aggregate behavior on static networks.** The left column shows results obtained for low density (*density=1*), the middle column has data for *density=3*, while the right column shows values for high density (*density=10*). The top row contains time series plotting aggregate difference between the choices versus time. The middle row shows the histogram, while the bottom row the box-plot of the same data. (The charts summarize a single example run of 10000 rounds for each case with N=100.)

### Binary Logit Rules on Erdős-Rényi Networks

In the following we will consider binary logit decision rules on Erdős-Rényi random graphs, where the single decision making parameter for the agents is social influence. (Erdős and Rényi, 1959) That is, the number of choices is fixed at *M*=2

(for example, car versus public transit) and the rules of $R$ are based on the probabilistic logit model as described below. (Ben-Akiva and Lerman 1985)

Given $M$ fixed at 2, we have two complementary alternatives. For convenience, we introduce a notation: for any alternative $j$ let $\bar{j}$ stand for the complementary choice. Moreover, let $U_{aj}^t \in [0,1]$ stand for the so-called "systematic" utility that agent $a$ associates with alternative $j$ at time $t$. Then, according to the binary logit model, the probability that agent $a$ chooses alternative $j$ is given by:

$$P_{aj}^t \equiv P_a^t(\psi_j|\Psi) = \frac{e^{(U_{aj}^t - U_{a\bar{j}}^t)}}{e^{(U_{aj}^t - U_{a\bar{j}}^t)} + 1}.$$

Following the spirit of (Aoki 1995) and (Brock and Durlauf 2001; Blume and Durlauf 2002) social dynamics is introduced by allowing the difference

$$U_{aj}^t - U_{a\bar{j}}^t$$

in systematic utility to be a linear-in-parameter $\beta$ function of the proportion $w_{aj}^t$ of agent $a$'s neighbors making each choice:

$$w_{aj}^t = \frac{\sum_{g \in G(a)} \chi(\sigma_g^t = \psi_j)}{|G(a)|} \text{ for } j \in [1,2],$$

where $\chi(\cdot)$ is the above membership function. For convenience, let's define $w^t$ as

$$w^t = w_{aj}^t - w_{a\bar{j}}^t.$$

Then the decision-making rules of the model take the following general form:

$$P_{aj}^t \equiv h_{aj}^t\left(\beta, w_{aj}^t, w_{a\bar{j}}^t\right) = \frac{e^{\beta \cdot w^t}}{e^{\beta \cdot w^t} + 1}.$$

The parameter $\beta$ indicates the level of certainty in the model. If it is fairly certain that the utility of alternative $\psi_1$ is greater than the utility of alternative $\psi_2$, then $\beta \gg 0$, and we have an effectively deterministic choice. If there is uncertainty as to which alternative has higher utility, then $\beta \sim 0$, and we have effectively a "fair coin toss" between the two alternatives:

$$P_{aj}^t = \frac{e^{\beta \cdot w^t}}{e^{\beta \cdot w^t} + 1} \sim 1, \text{ for } \beta \gg 0, \quad P_{aj}^t = \frac{e^0}{e^0 + 1} \sim 0.5, \text{ for } \beta \sim 0.$$

Initially, agents are assigned random choices drawn from an uniform distribution. They are also given a a set of neighbors, according to their position in graph $G$.

## Summary of Results on Static Networks

As seen earlier, for low values of the $\beta$ parameter the agents' decision making lacks any systemic property. Therefore, the aggregate outcome on static Erdős-Rényi networks is also essentially random: the difference between the number of agents choosing the two alternatives oscillates around 0. On the other hand, if $\beta$ is above a threshold, the agent population may converge to an unanimous decision (disturbed only by occasional random perturbations caused by the stochastic nature of the agent decision making process), depending on the network topology. In particular, if the network density (the uniform probability for each possible link to be present, expressed in our treatment as a multiplier of 1/N) is above the connectivity threshold (Molloy and Reed 1998), we observe unanimous aggregate outcomes, which may equally be either of the choices. Below this threshold, however, the difference between the sizes of the parties choosing the different alternatives oscillates around zero, just like in the low-$\beta$ case.(Dugundji and Gulyás 2005, 2006; Gulyás 2005) Fig. 1 demonstrates these observations with results from numerical experiments carried out using the RePast agent-based simulation package. (North, Collier and Vos 2006)[i]

On Watts-Strogatz networks we observe a similar two-phased behavior depending on the average distance between pairs of nodes in the network. (Watts and Strogatz 1998; Gulyás 2005)

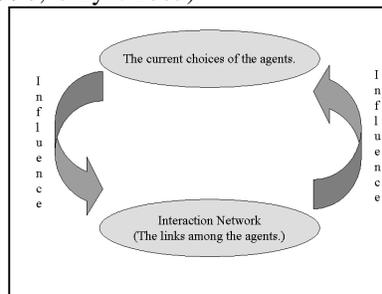

**Fig. 2: The circular influence pattern of endogenous network dynamics.**

# Social Network Dynamics

In the following we consider *dynamic* networks, i.e., cases when changes in the *interaction topology* (Δ) are at the same or a comparable time-scale than the repeated decision making of the agents. In principle, there are two approaches to incorporate network dynamics into our model. One regards changes in the social influence network as external to the process of making discrete choices. The other considers them to be inherently connected to the process of making discrete choices. In the former, exogenous case, changes can either be modeled as a stochastic process or, if more information is available on the network dynamics, a specific model of network evolution can be created. The focus of the present work is the latter case of endogenous network dynamics, where changes in the interaction topology are induced by the discrete choice dynamics ongoing on the very network. (See Fig. 2 for an illustration of this circular pattern of influence.)

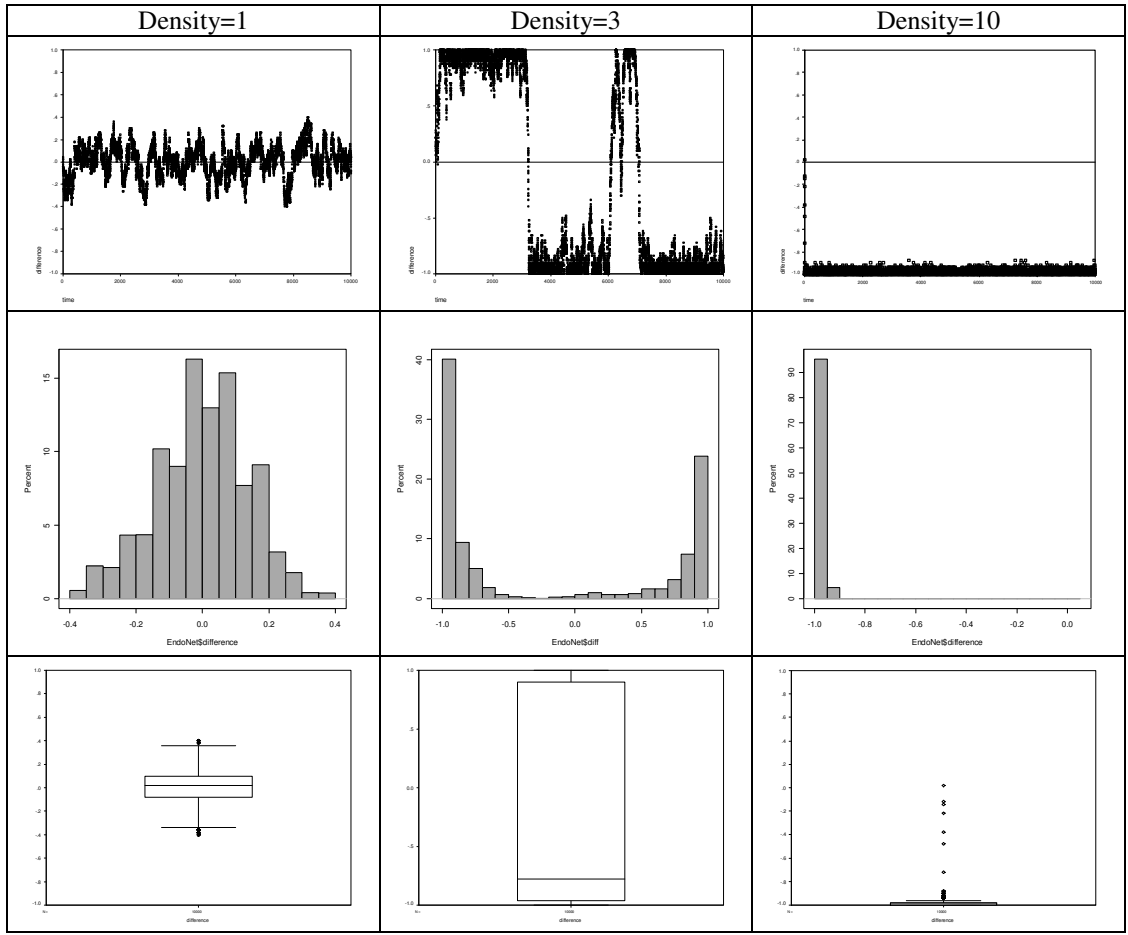

**Fig. 3: Density-dependent aggregate behavior on dynamic networks (*T*=0.5). The left column shows results obtained for low density (*density*=1), the middle column has data for *density*=3, while the right column shows values for high density (*density*=10). The top row contains time series plotting aggregate difference between the choices versus time. The middle row shows the histogram, while the bottom row the box-plot of the same data. (The charts summarize a single example run of 10000 rounds for each case with N=100.)**

# The Endogenous Network Model

The intuition behind our model of endogenous network dynamics is that a person's making a choice is a sign of her preference. Therefore, it can be interpreted as an increased likelihood that the person's future acquintances will have made the same choice. The same time, the person is less likely to maintain contact with former acquintances that made the opposite decision. We operationalize this approach as follows.

Let $d^t_i \in [0, N-1]$ denote the number of agent *i*'s neighbors that have made the same decision than agent *i* at time *t*. Similarly, let $z^t_i \in [0,1]$ denote the *ratio* of such agents and the total number of neighbors. (In the following we will omit the superscript *t* wherever it doesn't sacrifice understandability.)

Starting from a random network as in the static case, we define $\Delta d_i$, i.e., the *change* that occurs in the number of same-minded neighbors as a function of $z_i$. Initially, the agent will seek similar neighbors, but as the ratio reaches a threshold (*T*), it will rather opt for opposite-minded partners. Formally,

$$\Delta d_i = \begin{cases} L & \text{if } z_i \leq T \\ -L & \text{if } z_i > T \end{cases}$$

where *L* is a parameter (kept constant at 1 in our experiments). The above definition of $\Delta d_i$ is regarded as subject of the following constraints:

► Multiple links to the same agent and self-links (loops) are not allowed.
► The network density must be kept constant (as it is, from our earlier results, known to be a parameter to which aggregate behavior is sensitive). Therefore, each new neighbor 'costs' a link to the opposite group. If no such link exists, the new partner cannot be added.

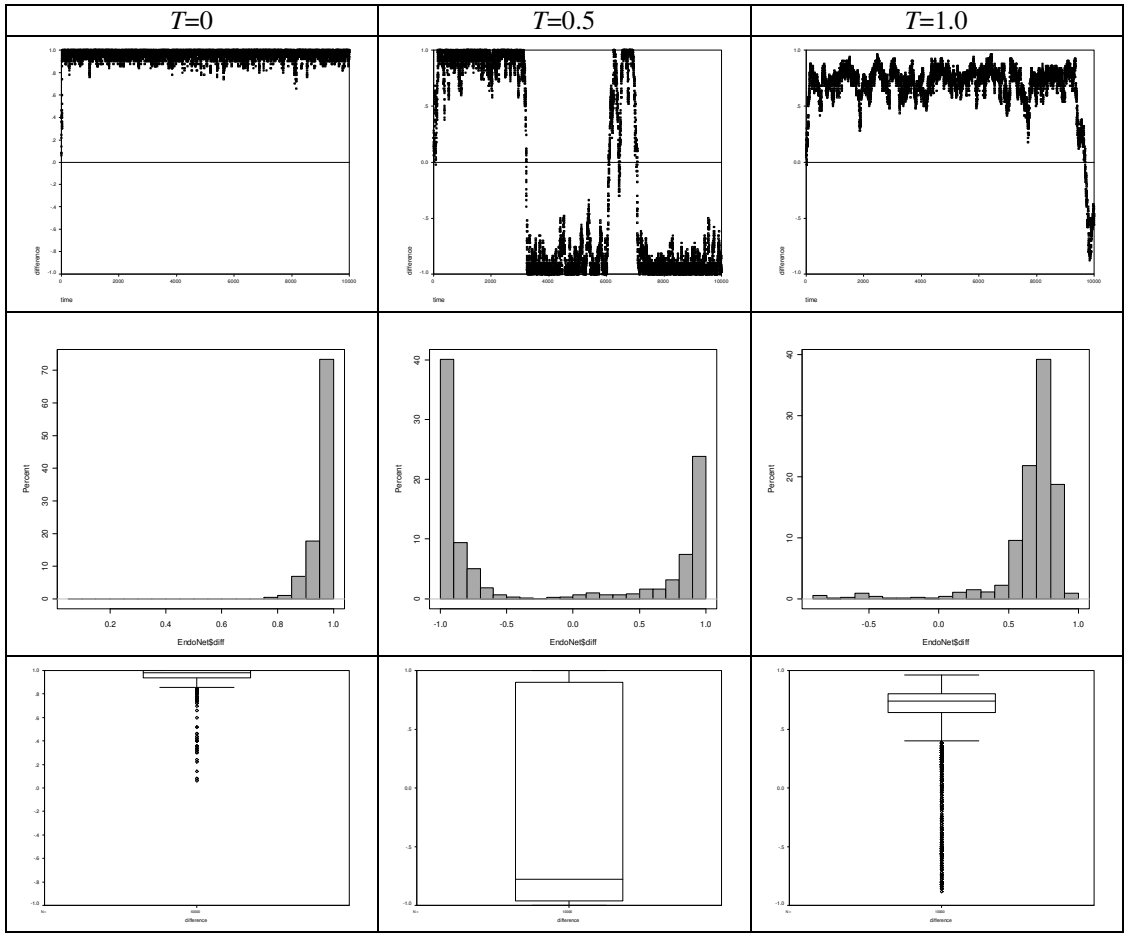

**Fig. 4: The aggregate behavior's dependence on the threshold parameter (*T*), for middle-density networks (*density*=3). The left column shows results obtained for *T*=0, the middle column has data for *T*=0.5, while the right column shows values with *T*=1. The top row contains time series plotting aggregate difference between the choices versus time. The middle row shows the histogram, while the bottom row the box-plot of the same data. (The charts summarize a single example run of 10000 rounds for each case with N=100.)**

Notice that the above definition identifies a *class* of potential 'future networks', among which we apply a probabilistic selection according to the following procedure. Agents update their partners at the end of each round sequentially in a randomized order. An agent first determines the number of links to be added and the choice of the new partners. Then it attempts to select the appropriate number of new partners randomly and adds them one by one as long as it can find an existing neighbor with the opposite choice to be dropped. The latters are also picked randomly. Notice that the pre-calculated number of new links ($\Delta d_i$) may only partially be achieved.

## Results

Our first observation is that, albeit the aggregate behavior remains dependent on the network density in endogenous networks as well, a completely novel type of behavior emerges for middle-density interaction topologies. As demonstrated on Fig. 3, the system shows the expected behavior for sparsely and densely connected networks. The aggregate difference between the choices oscillates around 0 in the former case, while it converges to almost 1 or -1 (i.e., each agent chooses one of the alternatives) in the latter. However, for a middle-range of densities the aggregate measure displays monumental swings. It first approaches one extreme (i.e., almost all agents opt for one alternative), but then, all of a sudden, the consensus breaks down, only to be quickly reached again – at the opposite end of the aggregate choice spectrum. The shifts are typically rather abrupt, albeit it is clear that the system goes through an almost continuous transition. That is, the new consensus is built around a few dissenters, whose influence manifests quickly.

This type of 'cycling' is completely missing from the dybamics observed on static networks, but is not unkown in the real-world. For example, many large cities went through similar cycles, when a large percentage of their population first moved out to the suburbs in the hope of a more peaceful life, only to migrate back once the commuting time was drastically increased by implied traffic jams. Naturally, it is not our intention to claim that our model offers an explanation for such real-world events. Yet, the latters' existence backs the feasibility of the observed behavior.

Fig. 4 analyses, in the middle-density range (density=3), how this cycling behavior depends on the threshold parameter ($T$) introduced in the definition of network dynamics. It is visible that positive feedback alone (i.e., when links to similar-minded agents are sought, $T=0$) is not enough for the occurrence of the phenomenon. On the other hand, pure negative feedback (i.e., when links to agents with the same opinion are constantly being dropped, $T=1$) does produce global shifts, albeit rather rarely. Also, the transition is less abrupt, and the convergence is to lower values of consensus. Nonetheless, 'cycling' is most pronounced when both positive and negative feedback is applied, e.g., for $T=0.5$.

## Related Works

A vast and growing body of literature is dedicated to networks, be them social, man-made or other real-world networks. Many of these works deal with abstract models that attempt to capture one or the other property of observed networks. See (Newman 2003) for a review. Similarly, many papers deal with opinion dynamics and their abstract models and the same is also true for discrete choices. (Hegselmann and Krause, 2002) (Ben-Akiva and Lerman 1985) (Brock and Durlauf 2001) Previous works from the authors connected the latter series of papers to those dealing with networks. (Gulyás and Dugundji 2003a, 2003b; Dugundji and Gulyás 2004a, 2004b, 2005, 2006)

On the other hand, only a few works have dealt with endogenous network dynamics. For example, (Pujol et al. 2005) and (Jin, Girvan and Newman 2001) introduce models of evolving networks, while (Takács and Janky 2006) apply a game theoretical approach to changing networks in the context of collective action.

However, we are not aware of any work that links discrete choice dynamics to endogenous network formation.

## Conclusions and Future Works

In this paper, we extended earlier results concerning discrete choices on networks by considering the endogenous dynamics of the underlying social networks. In particular, we studied a model that starts with interaction topologies belonging to the Erdős-Rényi class, where the behavior adopted by the agents feeds back to the underlying network structure. (E.g., a decision to commute by train instead of car increases the likelihood of train-choosing acquaintances.)

Our main observation is that the feedback-based endogenous network dynamics introduces a novel, third phase of aggregate behavior with monumental swings. For connected, but relatively low-density networks, the aggregate choice first approaches one extreme (i.e., almost all agents opt for one alternative), but then, all of a sudden, the consensus breaks down, only to be quickly reached again – at the opposite end of the aggregate choice spectrum.

This paper reported on the first set of results obtained via computational multi-agent based simulations. Future works include a more detailed analysis of our findings at greater length. In particular, we intend to analyze the size-dependency of the observed 'cycling' behavior. Moreover, we plan to perform an in-detail analysis of the mechanics of the abrupt shifts in the observed novel phase. A structural analysis of the networks emerging from the model is also planned. Finally, we intend to relax the initial condition that prescribes random initial networks that are unrelated to the, also random, initial choices of the agents.

# Acknowledgments

The partial support of of the Hungarian Government under the grant GVOP-3.2.2-2004.07-005/3.0 (ELTE Informatics Cooperative Research and Education Center) is gratefully acknowledged. This work started at the 2003 Santa Fe Graduate Workshop in Computational Economics. The authors would like to thank the organizers, Professors John H. Miller and Scott E. Page, for their support and fellow participants for their comments.

---

[i] In this paper only single runs are analyzed to keep our discussion compact. The selected runs are typical for their parameter combinations.